# Calculation of routing value in MPLS network according to traffic fractal properties

Tamara Radivilova
Infocommunication Engineering Department
Kharkiv National University of Radioelectronics
Kharkiv, Ukraine
tamara.radivilova@gmail.com

Lyudmyla Kirichenko
Applied Mathematics Department
Kharkiv National University of Radioelectronics
Kharkiv, Ukraine
lyudmyla.kirichenko@nure.ua

Oleksandra Yeremenko
Infocommunication Engineering Department
Kharkiv National University of Radioelectronics
Kharkiv, Ukraine
oleksandra.yeremenko.ua@ieee.org

*Abstract*— **Method for calculating routing cost of MPLS network is presented in the work. This method allows to minimize routing cost, taking into account traffic fractal properties, choice of traffic transmission path and quality of service requirements. The method uses values of the Hurst parameter and value of normalized spread of traffic values, which makes it possible to apply it to self-similar and multifractal data streams.**

*Keywords—Multiprotocol Label Switching, self-similar traffic, routing, calculation of routing cost*

## I. Introduction

Experimental and numerical studies conducted in recent decades show that traffic in many multimedia networks has self-similar (fractal) properties. Such traffic has a special structure continuing on many scales - in realization there is always a certain amount of very large bursts with a relatively low average traffic level. These bursts cause significant delays and packet loss, even when the total demand of all flows is far from the maximum allowable values. The reason for this effect lies in peculiarities of file distribution on servers, their sizes, in typical user behavior and it is largely related to changes in network resources and network topology. It turned out that initially non-self-similar properties of data streams, after processing on nodal servers and active network elements, begin to show pronounced signs of self-similarity. [1, 2]

At present, the Multiprotocol Label Switching Protocol (MPLS) is an effective solution to the problem of ensuring quality of network maintenance. Traffic management based on differentiated services in MPLS networks provides scalability of networks with several classes of services, resource reservation, fault tolerance and optimization of resource transfer. With the advent of MPLS-networks, they are studied in terms of impact of self-similar traffic properties on the quality of network maintenance. In works [1-3], studies of the effect of traffic self-similarity on the convergence of a real network in traffic control with different QoS (Quality of Service) were carried out. In [3], the multipath routing algorithm is proposed to reduce end-to-end delay and amount of losses. Models of fractal Brownian motion are considered as real traffic. Optimal number of data transmission paths for minimizing the average delay were calculated, and also optimal traffic distribution coefficient for multipath routing based on the game algorithm were calculated.

Currently, several approaches are being considered to reduce the effect of flows self-similarity in MPLS networks. In work [4] a mathematical model for estimating quality parameters and time for servicing network traffic was described. The data from model network were tested on a real multiservice data network, its parameters and properties were considered. In work [5], the effect of self-similar traffic and dynamic routing on the QoS of a multiservice network was analyzed. As input data an on-off model of self-similar traffic was used. The analysis was carried out for three classes of service of self-similar traffic. Also, relationship between different performance indicators and proposed traffic volume was investigated. In work [6], the XY distance routing algorithm (DPXY) is proposed to solve network latency problems caused by self-similar properties of traffic. Data was analyzed based on information about buffering and center distance when comparing the proposed algorithm with other routing algorithms.

Another approach, presented in [7], proposed a method for marking flows in FEC, taking into account the values of Hurst exponent and the dispersion coefficient. The definition of the burst level can also help in the marking of traffic flows, for example, in cases where higher priority traffic for a long time prevents the flow of lower priority streams. In [8], an approach is proposed to increase the amount of free bandwidth and the total cost of the network; when determining the path for traffic transmission and necessary bandwidth, self-similar traffic parameters are taken into account.

The purpose of this work is to improve the quality of service in MPLS network by preventing network congestion at peak traffic emissions with fractal properties. The paper presents a mathematical description of MPLS network operation, the properties of self-similar and multifractal traffic, and an algorithm is developed for calculating cost of routing, taking into account the fractal properties of traffic.

## II. MPLS NETWORK ARCHITECTURE

The MPLS network is conditionally divided into two functionally different areas: the core of network and edge area. In core, LSR (Label Switching Router) routers are only concerned with label switching. All FEC (Forwarding Equivalence Class) packet classification functions based on QoS (Quality of Service), as well as the implementation of additional filtering services, explicit routing, network load balancing and traffic management are taken over by LER (Label Edge Router).

Thus, the entire amount of intensive computing falls on edge area, and high-performance switching is performed in core, which allows efficient optimization of MPLS device configuration depending on their location in the network [8].

The architecture is based on MPLS label exchanging principle. On the edge of network MPLS label switching router LSR marks packets with special markings which determine further packet route to destination, indicating destination in label of each package, QoS information, packet path to destination, and other.

The process of label distribution between routers leads to establishment of MPLS paths on Label Switching Path (LSP). Each LSR router has a table in which old label value is replaced with a new one when processing a packet, after which the packet is sent to the next device by LSP path. The main problem is choosing the best LSP for traffic channel with QoS requirements in network. Within QoS, it is prudent to put traffic channels in priority in such way that QoS traffic is given priority [1-4, 7].

## III. DEVELOPMENT OF MATHEMATICAL MODEL OF MPLS NETWORK

The fact of multiplicity of routes is important, there are several alternative routes may exist for a route between two edge routers. The cost of routing and load balancing are proposed to take into account this feature of network construction.

In the mathematical model a network is represented as a graph $G = (V, E)$, where $V = \{1, 2, ..., N\}$ and $E = \{1, 2, ..., M\}$ is the set of routers and communication lines between them, respectively. The forward link $m$ has a capacity $u_m$ (in units/s). The entire set of nodes in MPLS network is divided into two subsets: $V^+ = \{V_i^+, i = 1, n_{LER}\}$ - LER, and $V^- = \{V_i^-, i = 1, n_{LSR}\}$ - LSR. When splitting into a subset, the following condition must be satisfied:

$$\begin{cases} V^+ \cup V^- = V \\ V^+ \cap V^- = 0 \end{cases}$$

Each element of a set $V^+$ can be both a source of traffic and a recipient. Traffic source means that this router receives traffic from an adjacent network (IP, MPLS, ATM, etc.) and it must be delivered to recipient node, which is also a contact point with adjacent networks. Lets consider the case where each edge router LER is both a source and a receiver. LSR can not be the recipient of traffic that has arrived to it from an adjacent network [8].

Suppose that at each moment $t \in T$ one of a routers receives traffic of intensity $\lambda(t)$ related to one of a service classes $q \in Q$ (in MPLS, 3 bits are allocated to the marking of service classes, thus 8 classes are possible). Each class corresponds to the values of the maximum allowable delay $\tau_q$ and the maximum permissible percentage of losses $l_q$.

It is assumed that set of QoS traffic channels indicated $Y$ (in one link $m$ may be a plurality of channels $y \in Y$), and that they have the following attributes:

- each traffic channel requires bandwidth $d_y$. These bandwidth requirements can be calculated from the contract with a client and / or from statistical data collected between input and output routers;

- QoS traffic routing performance is highly dependent on jitter, delay and reliability. Since traffic passes through LSP, latency at each transition will have a negative effect on efficiency of jitter and delay routing. In addition, when using fewer hops, the reliability of traffic transmission through channels increases, since a probability of failure LSP is reduced. Thus, traffic channels $y \in Y$ with QoS requirements $q$ have limitations on a number of hops in LSP. To implement this constraint, an allowed multiple paths are set $P_y = \left\{ p_y^1, ..., p_y^{L_y} \right\}$ for a path $L_y$ defined for each traffic channel.

Lets denote $\lambda_{P_y}^q(t)$ that at the moment $t$ the edge router receives traffic of intensity $\lambda$ related to the $q$-th class of service, which must be delivered to output router by any paths from the set $P_y$, not exceeding the specified maximum allowed delay values $\tau_q$ and maximum permissible percentage of losses $l_q$.

To describe the mechanism for releasing network resources that are used by traffic, when the transfer of this traffic is completed (this is based on data coming from the routing protocol supporting messages about available bandwidth, for example, CSPF), lets introduce some variables. A variable $\varepsilon_{P_y}^{q,t_0}(t) = \{0,1\}$ indicating that at a time $t$ traffic of class $q(\varepsilon = 1)$ ceased to flow to edge router, that was accepted for maintenance at a time $t_0$ and was to be transmitted along path of set $P_y$ to output edge router. This variable contains all the necessary data for determining the network resources for their release.

Each node of network $V$ at a moment $t$ is characterized by a performance $\mu$, the loss coefficient $X_V^q(t) \in X$, packet average waiting time in the queue $T_V^q(t) \in T$.

The variable $X_V^q(t) \in X$, $V = \overline{1, P_y}$ is equal to percentage of losses traffic on node with a class of service $q$ routed along path $P_y$ between edge routers at the time $t$. It is assumed that the probability of packet distortion in path can be neglected and losses occur exclusively at network nodes because of memory device overload. The following limitation is imposed on amount of losses for all network nodes $V = \overline{1, P_y}$:

$$0 \leq X_V^q(t), \sum_V^{P_y} X_V^q(t) \leq l_q . \quad (1)$$

Thus, it follows from the constraint (1) that total losses for traffic $\lambda_{P_y}^q(t)$ routed at a time $t$, should not exceed a maximum allowable value for a given class of service $l_q$. Losses are defined as a ratio of discarded data number to quantity received for maintenance. A value $X_V^q(t)$ is subject to minimization.

Limitations imposed on delay for all network nodes $V = \overline{1, P_y}$ are similar:

$$0 \leq T_V^q(t), \sum_V^{P_y} T_V^q(t) \leq \tau_q , \quad (2)$$

where $T_V^q(t)$ is an average packet latency of service class $q$ in queue at a node $V$ following a path $P_y$.

Execution of restriction (2) contributes to the fact that delivery time of packets does not exceed maximum allowable value for a given class of service $\tau_q$.

IV. SELF-SIMILAR AND MULTIFRACTAL DATA FLOWS PROPERTIES OF NETWORK TRAFFIC

It is known that traffic in multimedia networks has self-similarity properties. A stochastic process $X(t)$ is self-similar with a parameter $H$ if the process $a^{-H}X(at)$ is described by the same finite-dimensional distribution laws as $X(t)$. The parameter $H$, $0 < H < 1$ called the Hurst exponent, is the degree of self-similarity of the process. Along with this property, the parameter $H > 0.5$ characterizes the measure of the long-term dependence of the process, i.e. the decrease of the autocorrelation function $r(k)$ in accordance with the power law: $r(k) \sim k^{-\beta}$, $0 < \beta < 1$, $H = 1 - (\beta / 2)$.

If the stochastic process $X(t)$ has homogeneous fractal properties, which are determined by a single scaling Hurst parameter $H$, then it is called monofractal. In this case the initial moments of the process $X(t)$ can be described by the formula $\mathrm{E}\left[|X(t)|^q\right] \propto t^{Hq}$. Unlike monofractal processes, where all moments $\mathrm{E}\left[|X(t)|^q\right]$ show the same scaling, multifractal stochastic processes exhibit a more complex laws of scale behavior: $\mathrm{E}\left[|X(t)|^q\right] = \propto \cdot t^{qh(q)}$, where $h(q)$ is generalized Hurst exponent, which is a nonlinear function [9].

One of the reasons for the overload in multimedia networks is the presence of long-term dependence (persistence) in traffic: the closer the parameter $H$ to 1, the slower the correlations between the data decrease: $r(k) \sim \frac{1}{k^{2(1-H)}}$. In particular, this means that high data traffic is also likely to be followed by high data, that does not allow a buffer to be released quickly enough. One of the most important properties of network traffic is a presence of heavy tails of its one-dimensional distribution functions. The main property of a random variable with a heavy tail is that it exhibits high variability. In traffic, this is manifested by the number of bursts of large amplitudes at low intensity [10].

It was shown in [11,12] that a buffer queue length and a number of losses in network, when passing self-similar traffic, are determined by the Hurst exponent and the variance index, larger values of which correspond to larger scatter of traffic data. The results obtained in [13] show that queues and losses generated by multifractal traffic are determined by the Hurst exponent and the nonlinearity of the scaling function, which corresponds to traffic heterogeneity i.e. the scatter of traffic data. Thus, generalizing studies of self-similar and multifractal traffic, it can be argued that the main characteristics, which determine network overload are the Hurst exponent and the estimation of variation coefficient $S_v(\tau) = S / \overline{X}$, where $\tau$ is a time interval on which a passing traffic is fixed, $S$ is the standard deviation, and $\overline{X}$ is traffic intensity. The studies presented in [9, 13, 14] showed that for $H \geq 0,9$ or for persistent traffic with $S_v \geq 3$, the amount of data loss exceeds 5-10%.

V. METHOD OF CALCULATING A COST OF ROUTING IN MPLS NETWORK, TAKING INTO ACCOUNT TRAFFIC FRACTAL PROPERTIES

In MPLS architecture, it is possible to select routes based on individual flows, and different flows connecting a same pair of endpoints can follow different routes. In addition, if an overload occurs, routes that routed by MPLS architecture can be changed. The routing protocol, based on a channel state data, calculates a shortest paths (least cost routes) between input edge router and all others. In this paper, the method for calculating cost of routing, taking into account self-similar structure of traffic, which helps prevent network congestion during peak traffic bursts is proposed.

The value of the routing cost $c_m$ is assigned to communication link $m$ and may depend on several parameters, in particular speed, length, and reliability. The cost of the path $p_y^l$ is denoted $C_y^l$ and it equals to a sum of cost of communication lines: $C_y^l = \sum_{m \in p_y^l} c_m$. If $x_y^l(t)$ is bandwidth that is routed to allowable path $p_y^l$ of channel $y$ of traffic $\lambda_{P_y}^q(t)$, then:

$$\sum_{t \in T; l=1}^{L_y} x_y^l(t) = d_y, \quad \forall y \in Y, \quad \forall l \in \{1, ..., L_y\}.$$

The objective function, minimizing the cost of route on many paths $P_y = \{p_y^1, ..., p_y^{L_y}\}$, is as follows:

$$\sum_{y \in Y} \sum_{l=1}^{L_y} C_y^l x_y^l(t) \to \min. \qquad (3)$$

If a fractal traffic with large bursts is transmitted in network, it requires timely increase capacity of communication lines. To reflect the change in self-similar properties of flows, the costs of paths $C_y^l$ are updated at regular intervals and recalculated according to the formula

$$Cnew_y^l = \begin{cases} C_y^l, & H \leq 0.5; \\ C_y^l + (H - 0.5)C_0, & 0.5 < H < 0.9, S_v \leq 1; \\ C_y^l + (H - 0.5)(S_v - 1)C_0; & 0.5 < H < 0.9; 1 < S_v < 3; \\ C_y^l + C_0, & H \geq 0.9 \text{ or } H > 0.5, S_v \geq 3, \end{cases}$$

where $C_y^l = \sum_{m \in p_y^l} c_m$ is determined in accordance with the objective function (3), value $C_0$ is selected by network administrator considering network topology. The routing algorithm is not changed (path cost $Cnew_y^l = C_y^l$) if traffic has independent values ($H = 0.5$) or has antipersistent properties ($H < 0.5$). If $0.5 < H < 0.9$ and value and dispersion of data is small ($S_v \leq 1$), the value $C_y^l$ increases in proportion to value of the Hurst exponent. If value of the Hurst exponent $0.5 < H < 0.9$ and dispersion is large ($1 < S_v < 3$), the value $C_y^l$ increases in proportion to both characteristics. The cost with a maximum value $C_y^l + C_0$ is obtained at $H \geq 0.9$ or persistent traffic ($H > 0.5$) with the coefficient of variation $S_v \geq 3$. After recalculating the value of all paths the announcement of the state of paths is sent between routers.

## Conclusion

In the work the improved mathematical model of the MPLS network is proposed, which is presented in the form of a graph containing a set of routers and communication lines between them. All model parameters have a time dependence. The model allows to describe the behavior of MPLS network over time for different classes of incoming traffic, given the constraints on a waiting time of packet in a queue and a number of lost packets.

The method of estimating the cost of routing, which is based on the fractal properties of network traffic is proposed. Using the proposed method allows minimizing cost of routing, the choice of traffic transmission path and quality of service requirements. The developed method can be used to optimize network resources by simulation of network operation.